# Dynamics of pattern formation and emergence of swarming in *C. elegans*

## Authors

Esin Demir[1*], Y. Ilker Yaman[2*], and Askin Kocabas[1, 2, 3]†

[1]Bio-Medical Sciences and Engineering Program, Koç University, Sarıyer, Istanbul, Turkey 34450

[2]Department of Physics, Koç University, Sarıyer, Istanbul, Turkey 34450

[3]Koç University Surface Science and Technology Center, Koç University, Sarıyer, Istanbul, Turkey 34450

*These authors contributed equally.

†Corresponding author: akocabas@ku.edu.tr

## Abstract

Many animals in their natural habitat exhibit collective motion and form complex patterns to tackle environmental difficulties. Several physical and biological factors, such as animal motility, population densities, and chemical cues, play significant roles in this process. However, very little is known about how sensory information interplays with all these factors and controls the dynamics of collective response and pattern formation. Here, we use a model organism, *Caenorhabditis elegans,* to study the direct relation between oxygen sensing, pattern formation, and the emergence of swarming in active worm aggregates. We find that when thousands of animals gather on food, bacteria-mediated decrease in oxygen levels slowed down the animals and triggers motility-induced phase separation. Three coupled factors—bacterial accumulation, aerotaxis, and population density—act together and control the dynamics of pattern formation. Through several intermediate stages, aggregates converge to a large scale swarming phase and collectively move across the bacterial lawn. Additionally, our theoretical model captures behavioral differences resulting from the genetic variations and oxygen sensitivity. Altogether, our study provides many physical insights and a new platform for investigating the complex relationship between neural sensitivity, collective dynamics, and pattern formation.

## Introduction

Some animals have remarkable abilities to form complex patterns to cope with environmental challenges[1-9]. Formation of these biological patterns is particularly initiated by aggregation. At high population density, animals tend to aggregate as animal to animal interactions cause instability in uniform distribution[9-12]. Different factors such as motility of animals and chemical cues can trigger these instabilities by altering the behavior of animals. In particular, sensory information plays a primary role in controlling behaviors. Elucidating how environmental factors interplay with the sensory information is essential to improve our understanding of pattern formation in biological systems. However, this is a challenging problem in complex organisms.

Nematode *Caenorhabditis elegans* allows the use of sophisticated experimental tools and the results provide very precise information regarding the relation between genes, neural circuits, and behaviors[13-16]. This study is motivated by the possibility of developing a platform that combines both neuronal information and the collective dynamics in this model system.

*C. elegans* shows a variety of collective responses including, social feeding[17-20], starvation-induced clustering[21], hydrodynamic networking[22] and collective swimming[23,24]. Recent studies have shed light on some physical[22,25,26] and biological[27-29] aspects of these behaviors and their variations. Oxygen sensitivity is particularly the main source of the variation. Laboratory strain (N2) is known to be solitary nematode with a broad range of oxygen preference while natural isolates exhibit strong aggregation and sharp aerotaxis[27,30]. Further, oxygen sensing, feeding and aggregation behavior are all intricately related in *C. elegans*. Bacteria utilize oxygen for growth and *C. elegans* seeks low oxygen levels for locating the bacteria. However, in a dense population, all these factors become extremely coupled[31-33]. In this study, we aimed to develop a general model dealing with the dynamics of all these factors.

Here, we report that, *C. elegans* including the laboratory strain (N2), forms complex patterns during feeding. When thousands of worms are forced to feed together, aggregation-induced bacterial accumulation and oxygen depletion create unstable conditions and further trigger phase separations. The principle of phase separation is mainly based on sudden change in the animal's motility[11,34]. We also found that the dynamics of the entire process is controlled by the sensitivity of the oxygen-sensing neurons, which gives rise to strong variations in animals' collective response. Finally, we also observe the convergence of self-organized patterns to the motile swarming body as a result of complex interactions between oxygen diffusion, bacterial consumption, and motility of animals.

## Results

To investigate the collective response of *C. elegans,* we began by imaging animals during food search in a crowded environment. We collected thousands of animals (N2) in a droplet and let it dry on an agar plate. The droplet was put a few centimeters away from food. After drying, worms spread and searched for food by tracing the attractive chemicals. This experimental procedure allowed us to increase the worm density around a bacterial lawn to observe their collective movement. When animals found the bacteria, they usually slowed down and penetrated the lawn.

However, we observed that, at high densities the majority of the worms did not penetrate the lawn. After reaching the lawn, they suddenly stopped and started accumulating (Fig. 1a). As the worms accumulated, they formed a huge aggregate which covered the entrance. Eventually, this large aggregate gained motility and swarmed across the lawn (Fig. 1a, 1b, 1c, Supplementary Video 1). This observation was quite different from the previously reported response of the reference strain N2 which generally behaves solitarily[17,32]. Further, we tested different mutant strains deficient in several sensing mechanisms and essential neuromodulators; we observed similar aggregation and swarming response in a wide range of animals (Supplementary Figure 1). We also used various types of food sources including *B. Subtilis* biofilm, filamentous bacteria, or extremely thick bacterial lawns. Surprisingly, animals showed very rich collective response ranging from complex pattern formations to large scale swarming (Fig. 1d, 1e, 1f, Supplementary Figure 2). Altogether, these results suggest that the conditions observed in a dense population can broadly trigger the collective response in *C. elegans*.

This collective response only occurs on a bacterial lawn. Previous studies reported the same conditions for aggregating strains[17,32]. However, how the presence of bacteria contributes to this process is unclear. To clarify this point, we used GFP expressing bacteria (OP50 GFP) to follow the entire dynamics. Time-lapse fluorescence microscopy revealed that the bacteria are concentrated within the aggregates. This bacterial accumulation further promotes the collective behavior of animals (Fig. 1f, SupplementaryFigure 3, Supplementary Video 2, 3). Following the accumulation of bacteria, the motility of animals is strongly suppressed. The naïve hypothesis explaining this observation is that the process is mainly based on the capillary meniscus around the animals[35,36]. When animals form aggregates, the structure becomes porous. Thus, due to the capillary effect, the porous aggregate can hold more bacterial suspension. Eventually, concentrated bacteria could make the formation of aggregate more favorable by conditioning the oxygen levels. We can conclude that concentrated bacteria is the primary factor triggering the formation and the maintenance of the aggregation.

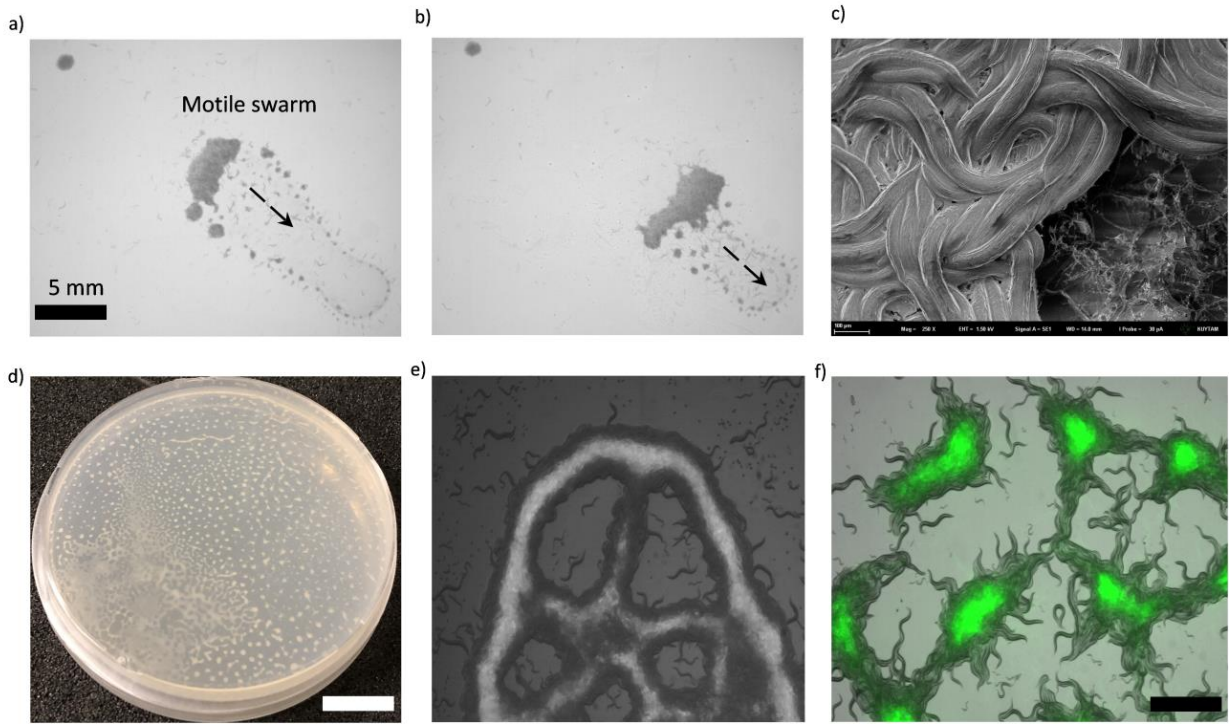

*Figure 1: Collective response and pattern formations in C. elegans. a, b Thousands of animals collectively swarm on a bacterial lawn. a While searching for food, worms aggregate around the edge of the bacterial lawn. Small aggregates merge and form a massive swarm. b Swarming body eventually gains motility and moves across the lawn while consuming the bacteria. c A cryo-SEM image of the swarming body. The heads and the tails of the animals are buried, and the bodies extend on the surface of the swarm. d, f Sample images of Turing-like patterns formed by animals on an agar surface under various conditions, such as starvation-induced and on a biofilm. The magnified version of the image is given in the supplementary information. f Fluorescent image of worms aggregating near the GFP-labeled bacteria. Due to the capillary effect and diffusion of bacteria, worms concentrate the bacteria within the aggregates.*

We then focused on the effects of oxygen on the entire process. Previous studies have already shown that the bacterial lawn can decrease oxygen concentration $[O_2]$[27,32]. To directly observe this effect, we used fiber optic sensors to measure $[O_2]$ in both bacterial suspension and in the swarming body. We observed that the oxygen level drops below 0.1% in the bacterial suspension

(Supplementary Figure 4). Moreover, we found similar depletion in [$O_2$] in the swarming body (Fig. 2a, Supplementary Video 4). These results imply that concentrated bacteria deplete [$O_2$] in the swarming body. It should be noted that the fiber optic sensor can only be used to record the oxygen levels in the swarm liquid. Although [$O_2$] is very low, the animals' bodies can extend on the surface of the swarm and they can get sufficient $O_2$ from the environment. Besides the ambient environment, wet agar surface can also provide oxygen by lateral surface diffusion. We independently validated these contributions by covering the aggregate with a glass slide and by using an agar chunk containing oxygen depleting chemical $Na_2SO_3$, (Supplementary Figure 5, Supplementary Video 5 and 6). Moreover, we observed that $Na_2SO_3$ based oxygen depletion on the agar surface could initiate the aggregation behavior in the bacteria-free region (Supplementary Video 7). From these results we can conclude that animals effectively experience the average $O_2$ levels defined by the ambient environment and the swarm liquid. Altogether, ambient oxygen levels and oxygen depletion are essential factors for aggregation.

To quantify the effects of [$O_2$] on the motility of animals, we measured the response of the animals under changing oxygen concentrations. Figure 2b compares the velocity profiles V(O) of both, aggregating strain *npr-1* (DA609) and solitary strain N2 and the velocity of the animals moving in the bacteria-free region (off-food). Both strains responded to very low [$O_2$] by increasing their velocities, which mainly triggers dispersive behavior (Supplementary Video 8). On the contrary, *npr-1* suppressed its motility around intermediate oxygen levels (7-10%)and showed a sharp response when oxygen levels exceeded 15%. In contrast, N2 suppressed its motility in more broader range of oxygen levels, but their velocity slowly increased as [$O_2$] reached 21% or more. These differences appear to be originating from the sensitivity of the oxygen-sensing neurons URX that shape the overall oxygen preferences of the animals[26,28,37]. Although both *npr-1* and N2 perform aerotaxis at 7-10% oxygen, N2 show weaker aerotactic response, thus, they have broader range of oxygen preference. All these critical features of the strains can be extracted from oxygen-dependent response curve V(O) which is directly related to neuronal sensitivity.

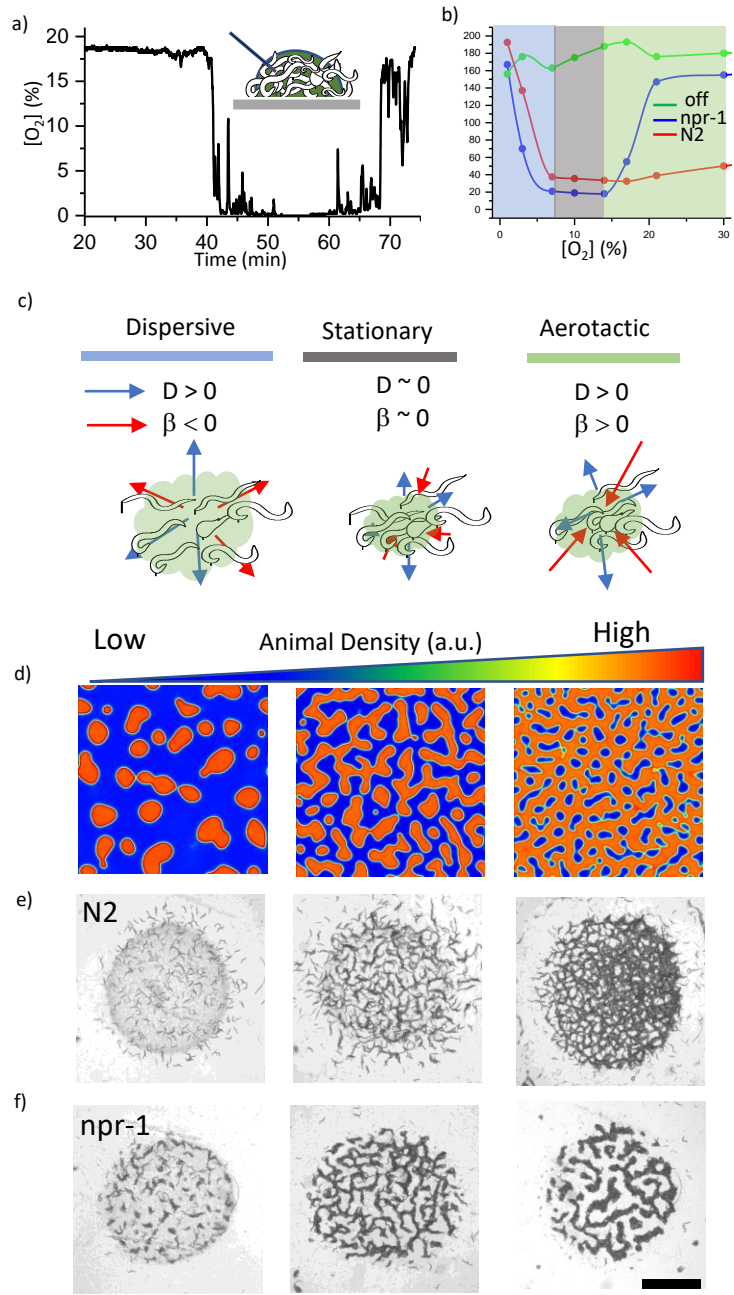

*Figure 2: Oxygen sensitivity and the dynamics of pattern formation. **a** Accumulation of bacterial suspension depletes oxygen in worm aggregates. Oxygen levels measured by the fiber optic sensor while the swarm passes through the sensor. **b** Oxygen-dependent motility of animals. The slope of*

*the curve defines the aerotactic response of the animals. The comparison of velocity profiles of N2 (red), npr-1(blue), and off-food (green) response shows the difference between individual worms. Colored regions indicate the dispersive, stationary and aerotactic phases. **c** Schematic representation of each phase. The competition between animal dispersion and aerotactic response defines the collective behavior of worms. **d** Simulation results of the model at steady-state with various animal densities. The pattern formation strongly depends on worm density. At low density, worms create clusters. As the worm density increases, clusters converge to stripes and hole patterns. **e, f** Experimental results of density-dependent pattern formation. **e** At low worm densities (500 worms/cm$^2$), N2 worms do not form patterns. As the density increases, we observed stripes (2000 worms/cm$^2$) and holes (6000 worms/cm$^2$). **f** The strong aerotactic response of npr-1 strain enables worms to form patterns even at low densities. Increase in the worm density causes change in the instability conditions and tunes the patterns.*

Motility suppression behavior has been observed in a variety of organisms ranging from bacteria to mussels[34,38-43]. Generally, animals tend to slow down when they come together. The entire process is represented by the density dependence of the animal movement which can lead to the formation of patterns. However, in *C. elegans* we observed indirect density-dependent suppression. Without bacteria, animals move fast (Fig. 2b). In striking contrast, the presence of bacteria results in oxygen depletion and motility suppression. These findings suggest that in the dynamics of oxygen, bacterial concentration and animal motility are the essential physical factors controlling collective behaviors of *C. elegans*.

To gain a more quantitative representation of pattern formation, we developed a mathematical model. We followed the notation and the framework developed for active chemotactic particles[44]. We set two separate differential equations to represent worm density (W) and oxygen kinematics (O) in two-dimensional space. V(O) is the oxygen-dependent motility response of individual animals. This factor served as an experimentally measurable sensory curve of the animals. This curve defines competition between diffusivity and aerotactic motility. Both effects became minimal around the optimum oxygen level (7–10 %) at which the animals were almost in a stationary phase (Fig. 2b). In low oxygen region, dispersion and reversal of aerotaxis promoted motility. On the other hand, at high oxygen levels, aerotaxis dominated the dynamics and promoted aggregation (Fig. 2c). These are the Keller-Segel like equations[45,46] with both, nonlinear diffusion and chemotactic sensitivity;

$$\frac{\partial W}{\partial t} = \nabla[D_W \nabla W] + \nabla[\beta W \nabla O]$$

$$\frac{\partial O}{\partial t} = D_O \nabla^2 O + f(O_{am} - O) - k_c W$$

Here, $D_W = \frac{V^2}{2\tau}$ represents motility-dependent dispersion of the worms, $D_O$ is the diffusion coefficient of oxygen on the surface, $\beta = \frac{V}{2\tau}\frac{\partial V}{\partial O}$ is the aerotactic coupling coefficient indicating the strength of the animal response to oxygen gradient. $f$ is the penetration rate of the oxygen from the air to the water surface, $O_{am}$ is the ambient oxygen level, and $k_c$ is the rate of oxygen consumption by worms and bacteria. Here, we assume that bacterial density is linearly proportional to worm density. Furthermore, the time scale of the bacterial consumption is much longer than the aggregation time, thus we ignore the bacterial consumption term in the equations.

First, we focused on the instability criteria. Following the linear stability analysis of the solution in the uniform phase, we predicted that instability occurs when $W_{eq}\beta k_c > fD_W$. The detailed derivations are given in the supplementary information. This criterion suggests that population density and aerotactic response favor instability; however, dispersion of animals opposes it. To verify these conditions, we performed simulations by tuning initial worm density. Depending on the density, we observed the formation of uniform distribution, aggregates, stripes, and holes (Fig. 2d). These are the basic patterns that are frequently seen in many biological systems[11,47].

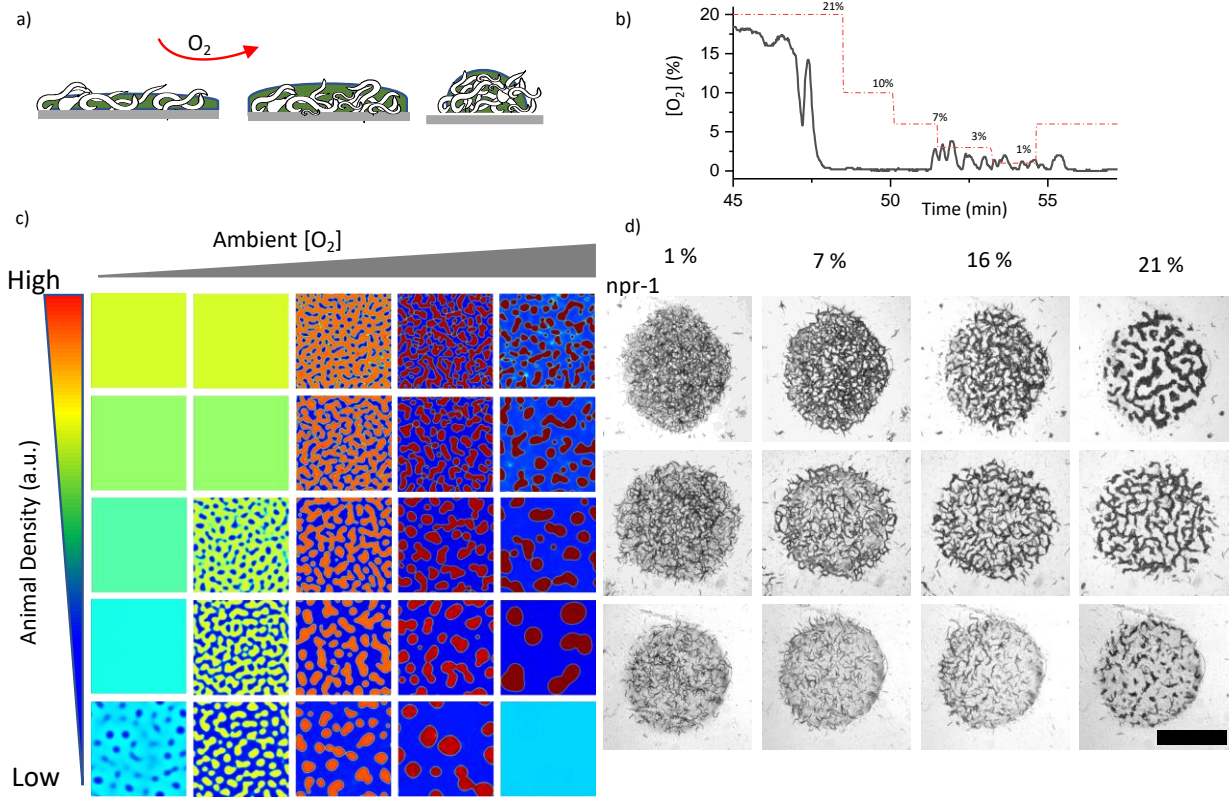

*Figure 3: Combinatorial effect of ambient oxygen and worm density. a Schematics show the response of worm aggregates to changing ambient oxygen levels. As oxygen level increases, the size of the clusters shrink and form circular aggregates which balance the oxygen penetration and diffusion. b Measured oxygen concentration in worm aggregates. The dashed red line shows the ambient oxygen level. c, d Simulation and experimental results of pattern formation under various worm densities and ambient oxygen levels. Patterns are formed when the instability criterion is satisfied. The instability zone is bounded by the uniform stable population.*

Our simulations also predict that under low aerotactic response, the increase in population density could complement the instability criteria and give rise to pattern formation. This is because the multiplication of density and aerotactic response simply shifts the dynamics to a new instability zone. This prediction could explain the response of solitary N2 in a dense population. N2 shows weak aerotactic response which results in broad oxygen preference and patterns only appear in a dense environment. On the other hand, the strain *npr-1* with sharp aerotactic response can form similar patterns even at low population density. We tested this hypothesis experimentally (Fig. 2e, 2f, Supplementary Video 9, 10) and found that N2 forms stripes and hole-like patterns only at high population density, whereas *npr-1* could form small aggregation patterns even at a low density.

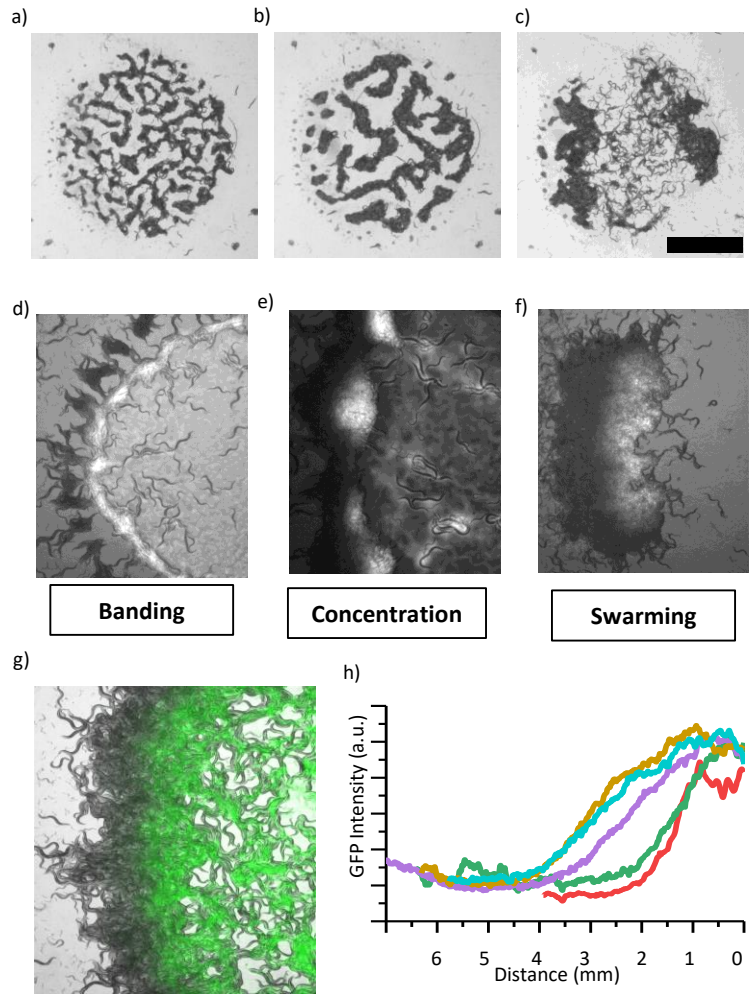

*Figure 4: Coarsening event and the emergence of swarming behavior. a, b, c After the formation of patterns a animals locally consume bacteria which eventually increases the motility of animals. Further, patterns merge and coarsen. b Consumption of food eventually results in the formation of a large aggregate which resembles the swarming body. d, e, f Different stages of a swarming response observed during food search behavior in a dense population. d Worms accumulate around the lawn. Unlike isolated worms, diffused bacteria cause the worms to slow down and then the worms form a band around the edge of lawn which eventually converges to aggregates. e Small aggregates concentrate bacteria and further accumulate the incoming worms. Consumption and diffusion of bacteria from the lawn leads to a bacterial gradient within the aggregate. f Following the formation of a gradient, large aggregate gains motility. The swarm moves through the lawn until finishing the entire lawn and then dissociates. g Superposition image of swarming animals*

*and GFP-labeled bacteria indicating the gradient formation across the swarm.* **h** *A gradient profile gradually extends into the swarm.*

Ambient oxygen is the second experimentally controllable factor driving the dynamics of the system. The onset of the pattern formation can be controlled by tuning the ambient oxygen levels at a fixed population density. Further, we tested the contribution of ambient oxygen levels to pattern formation. To be able to control the $[O_2]$, we performed the experiments in a chamber where the flow of $O_2/N_2$ mixture is precisely controlled (Supplementary Figure 6). First, we measured the $[O_2]$ in a large aggregate accumulated around the bacterial lawn. We started the experiment at 21% $O_2$ and sequentially decreased the $[O_2]$ to 1%. We observed that the aggregates response to decrease in $O_2$ level by increasing their surface area (Fig. 3a, Supplementary Figure 7, Supplementary Video 11). However, $[O_2]$ in the aggregate remained low until it reached 7% (Fig. 3b). Below this level, the pattern became very porous and oxygen would directly leak into the fiber sensor which led to small fluctuations in readings. It should be noted that the $[O_2]$ at 1% completely randomized the patterns by increasing motility and reversing aerotaxis. Subsequently, we increased the $[O_2]$ which resulted in similar stripe and hole patterns across the region (Supplementary Figure 8, Fig. 3c). We noticed that patterns formed by N2 strain did not change significantly during oxygen scans (Supplementary Figure 9). This was expected, because N2 exhibits low and almost flat aerotactic response within the same range of oxygen levels.

We further explored the combinatorial effects of oxygen and worm density on pattern formation. As predicted by the instability criterion, in a 2-dimensional parameter space, the onset of dot, stripe and hole patterns is observed around the zone bounded by high and low uniform animal densities (Fig. 3d, 3e). Increase in worm density or decrease in ambient oxygen levels transforms the dot-shaped structure to stripe and hole patterns. The main intuition behind the pattern evolution can be captured by analyzing the oxygen kinematics. Aggregates with circular shapes minimize the perimeter and decrease the lateral surface diffusion of the oxygen from the side. However, elongated shapes can increase the perimeter and oxygen diffusion in a fixed surface area.

The other interesting feature of the pattern formation is the coarsening event. We sought to know how the shape of the patterns evolves in time at a fixed oxygen level. Our time-lapse imaging validates the coarsening (Fig. 4a-c). Interestingly, in later stages, patterns merge and form very large aggregates. This type of coarsening shares visual similarities with our initial swarming experiments (Fig. 1a, 1b). We noticed that in both cases, the consumption of bacteria is significant and may cause an additional impact on the collective dynamics. The similar effect of bacterial depletion was also proposed to explain the motion of the aggregating strain npr-1[25]. To investigate the details, we measured the bacterial concentration. GFP signal revealed different bacterial concentrations across the swarm; the front edge of the swarm has more bacteria than the back (Fig. 4d-f). Worms in the swarm consumed bacteria and the food continually diffused from the front edge towards the back. As the swarm grows, the gradient profile gradually extends into the swarming body with the average decay length of around 3–4 mm (Fig. 4g, 4h).

Next, we tested whether this concentration profile could change the activity of the animals. The activity of the animals increased towards the back which suggests that the animals crawling at the front edge encountered more bacteria than those at the back. To quantify the activity profile, we measured the mean velocity of animals using Particle Image Velocimetry (PIV) analysis. Indeed, the velocity increased towards the back (Supplementary Figure 10). This response is consistent with our V(O) curve where the animals perform off-food response. Without availability of bacteria, animals start moving fast. On the whole, due to the balance between bacterial consumption and diffusion, swarm body gains motility and moves across the bacterial lawn (Supplementary Figure 11).

## Discussion

The dynamics of pattern formation in biological systems depend on many intricately related factors. The theory of pattern formation in active particles provides a powerful framework to explain the complex interactions between these factors[38,44]. Using Keller-Segel model[45,46] and motility-induced phase separation principles[11,34], our study throws light upon new physical and biological insights of this complex dynamics. Our results revealed four essential factors of pattern formation. First, hydrodynamic interactions between worms initiate the process of bacterial accumulation within the aggregates, which is the first, and a critical step in the pattern formation. Second, oxygen dependent motility of the animals controls the competition between aerotaxis and animal dispersion. This competition links the neuronal sensitivity to the collective response of the animals. Third, the population density can compensate the neuronal sensitivity to convert the behavior of solitary animals to aggregation. Finally, a gradient profile is formed across the aggregate due to the consumption of bacteria, which leads to the initiation of forward motility and swarming behavior. Altogether, experimental results and mathematical models of this study may lead to future studies, which will further aid us in understanding the complex dynamics of biological systems and in designing a new generation of collective robots[48,49].

## Methods

### *C. elegans* strains

Strains were grown and maintained under standard conditions unless indicated otherwise. All the strains were obtained from Caenorhabditis Genetics Center (CGC).

Strain List: *npr1* (DA609), *npr1* (CX4148), *gcy35* (AX1295), *cat2* (CB1112), *tph1* (MT15434), *bas1* (MT7988), *bas1 & cat4* (MT7983), *tax2* (PR694), *tax4* (PR678), *tax2 & tax4* (BR5514), *nsy1*

(AWC off killed-CX4998), *eat2* (DA465), *mec4* (CB1611), *mec10* (CB1515), *osm9* (VC1262), *trpa1* (RB1052), *dig1* (MT2840), *trp4* (VC1141), *lite1* (*ce314*), *daf19 & daf12* (JT6924).

**Swarming protocol**

Nematode growth media (NGM) plates having diameter of 9 cm were used for maintaining the worms. NGM plates were seeded with 1 ml of OP50 culture. After the worms consumed the bacteria, three NGM plates were washed with 1× M9 buffer and the wash was centrifuged twice at 2000 rpm for 30 seconds. Centrifugation was repeated (up to six times) until a clear supernatant was obtained. Since multiple centrifugation cycles might affect the activity of worms, the cycles subsequent to the first two cycles were carried out for 10 seconds at 2000 rpm. After cleaning the worms, a 150 µl l worm droplet was put on a 6 cm plate seeded with 100 µl (0.5 × 1 cm) of OP50 culture. The swarming pattern of the worms was observed from day one to day twelve of the experiment. The optimum swarming pattern was observed in six days old plates and hence six days old plates were used in all further experiments unless indicated otherwise. For NGM plates, M9 buffer preparation and worm synchronization was done based on the standard protocols given in the wormbook.

**Effect of Oxygen levels on Swarming**

A plexiglass chamber was used to control the ambient [$O_2$] (Supplementary Figure 6). A 50 sccm mixture of $O_2$ and $N_2$ was used and flow rates were controlled by a flow controller. [$O_2$] was measured by using a normal oxygen sensor (PreSens, Microx TX3). At the beginning of the swarming experiments, the oxygen levels in the chamber were adjusted to 21%. After swarming was first observed, oxygen levels were sequentially decreased to 10%, 7%, 3%, and 1% at an interval of ten minutes. Then, oxygen levels were increased in reverse order. The experiments were recorded using Thorlabs DCC1545M CMOS camera with a Navitar 7000 TV zoom lens.

## Acknowledgments

This work was supported by an EMBO installation grant (IG 3275); BAGEP young investigator award and TUBITAK (project no115F072 and 115S666). We thank H. Kavakli, M. Iskin, F. Balci, P. A. Ramey for discussions and comments.

## Author contributions

E.D. observed the swarming response and E.D. and A.K. performed swarming experiments and mutant screen. Y.I.Y. and A.K. performed the pattern formation experiments, designed the imaging systems, developed the mathematical model, performed the simulations and carried out image processing analysis of the experiments. E.D., Y.I.Y. and A.K. wrote the manuscript.

## Competing interests

Authors declare no competing interests.